\documentclass[aps,twocolumn,twoside]{revtex4}
\usepackage{epsfig,amsmath,amssymb,dblfloatfix,bm}
\usepackage{color}

\newcommand{\ba}{\begin{eqnarray}}
\newcommand{\ea}{\end{eqnarray}}
\newcommand{\bmath}{\begin{mathletters}}
\newcommand{\emath}{\end{mathletters}}
\newcommand{\ban}{\begin{eqnarray*}}
\newcommand{\ean}{\end{eqnarray*}}

\newcommand{\bsub}{\begin{subequations}}
\newcommand{\esub}{\end{subequations}}

\begin{document}

\title{Critical excited states in quantum metrology}
\author{Marek Kucha{\v r}}\email{marek.kuchar@isibrno.cz} 
\affiliation{Czech Academy of Sciences, Institute of Scientific Instruments, Kr\'alovopolsk\'a 147, 61264 Brno, Czech Republic}
\affiliation{Charles University, Faculty of Mathematics and Physics, Ke Karlovu 2, 12000, Prague, Czech Republic}

\author{Michal Macek}\email{michal.macek@isibrno.cz} 
\affiliation{Czech Academy of Sciences, Institute of Scientific Instruments, Kr\'alovopolsk\'a 147, 61264 Brno, Czech Republic}

\begin{abstract}
We investigate criticality-enhanced sensing near ground- and excited-state quantum phase transitions in a strongly coupled qubit-boson system. Accounting for finite-size effects, we design driving protocols that exploit diverging susceptibilities to enhance signal-to-noise ratios in parameter estimation using experimentally relevant generators. We show that the excited-state criticality provides a metrological 
advantage through single-mode bosonic squeezing in regimes that circumvent the 'death of adiabaticity', 
limiting conventional critical quantum metrology with ground-state criticality.
\end{abstract}

\date{\today}

\maketitle

{\it Introduction—}The quest for quantum-enhanced precision has motivated the search for states with large quantum Fisher information (QFI), which sets the ultimate limits in quantum parameter estimation \cite{braunstein1994statistical,giovannetti2004quantum,paris2009quantum}. QFI is directly related to susceptibilities to small parameter perturbations, establishing a natural link with quantum phase transitions (QPTs) \cite{CarrBook,SachdevBook,Frerot18}. This connection has been explored near the ground-state QPTs \cite{Wang14,Wei18,Hwang15,Pedernales16,Duan21,Garbe20}, where divergent susceptibilities and enhanced correlations can yield metrological gain. However, the closing of the energy gap at a critical point leads to divergent characteristic timescales, severely limiting adiabatic preparation of high-fidelity ground states in realistic protocols \cite{Mihailescu26,Gietka26,Liu21,Chu21}. 

The critical behavior of many systems is not restricted to the ground state, as excited-state quantum phase transitions (ESQPTs) \cite{Cejnar21,Monodromy,Caprio8,Corps25} arise at specific points of the spectrum. They are characterized by non-analycities in the density of states \cite{Stransky14,Stransky16} and abrupt structural changes in eigenstates at critical energies, often associated with unstable stationary points in the underlying semiclassical phase space. Despite extensive studies of their dynamical \cite{Santos15, Bernal17} and statistical \cite{Cejnar08,Zakaria21,Bernal21} signatures, the potential of ESQPTs as a resource for quantum metrology remains largely unexplored, with existing studies primarily using their connection to non-equilibrium dynamical criticality \cite{Guan21,Zhou23,Mumford26}. In this work, we show that unstable fixed points associated with particular ESQPTs can be directly exploited for quantum metrology. Using the extended quantum Rabi model (ERM) \cite{Liu17,Stransky21} as a paradigmatic single-particle platform, we demonstrate that eigenstates near the ESQPT critical energy exhibit markedly enhanced QFI with respect to experimentally relevant generators. Moreover, their preparation can circumvent the restrictions imposed by the vanishing energy-gap at ground-state critical points, with the consequent breakdown of adiabaticity.

{\it Model}—The qubit-energy normalized ERM, representing a fundamental single-boson interaction between a two-level system and a bosonic mode, has the following form in the two-level basis \cite{Kuchar26}:
\begin{gather}\label{1}
\hat{h} =
  \begin{bmatrix}
   \frac{1}{\Delta}\hat{a}^\dagger\hat{a}-\frac{1}{2} &
   \frac{\lambda}{\sqrt{\Delta}}\left(\frac{1+\delta}{2}\hat{a}+\frac{1-\delta}{2}\hat{a}^\dagger\right) \\
   \frac{\lambda}{\sqrt{\Delta}}\left(\frac{1-\delta}{2}\hat{a}+\frac{1+\delta}{2}\hat{a}^\dagger\right) &
   \frac{1}{\Delta}\hat{a}^\dagger\hat{a}+\frac{1}{2} 
   \end{bmatrix}\,.
\end{gather}
\noindent The frequency ratio $\Delta$ of the qubit to the bosonic-mode was shown \cite{Hwang15, Stransky21} to be the inverse of an effective Planck constant representing the system size. Quantum criticalities of the system therefore occur in the thermodynamic limit $\Delta\to\infty$. Interaction strength $\lambda\geq0$ and interaction weight $\delta\in\langle-1,1\rangle$ span the parameter space upon which the following phases \cite{Stransky21,Kuchar26} can be found:
\begin{itemize}
    \item \textit{Normal phase} (\textbf{N}): Occurs for arbitrary $\delta$ in the $\lambda<1$ region. The qubit and phonon degrees of freedom are weakly coupled. The semiclassical Hamiltonian has just a global minimum (see Fig. 1(a)).
    \item \textit{The first superradiant phase} (\textbf{S1}): Spans a connected intermediate $\lambda\in(1,|\delta|^{-1})$ region. The semiclassical Hamiltonian develops a double-well structure in the $x$-direction and a saddle-point in the origin (see Fig. 1).
    \item \textit{The second superradiant phase} (\textbf{S2}): Covers the remaining disjoint $\lambda>|\delta|^{-1}$ region. The double-well structure remains in place, but the separatrix moves down in energy and splits the single saddle-point into two along the $p$-direction, leaving a local maximum at the origin (see Fig. 1(b)).
\end{itemize}

\begin{figure}[t!]
  \centering
  \includegraphics[width=\linewidth]{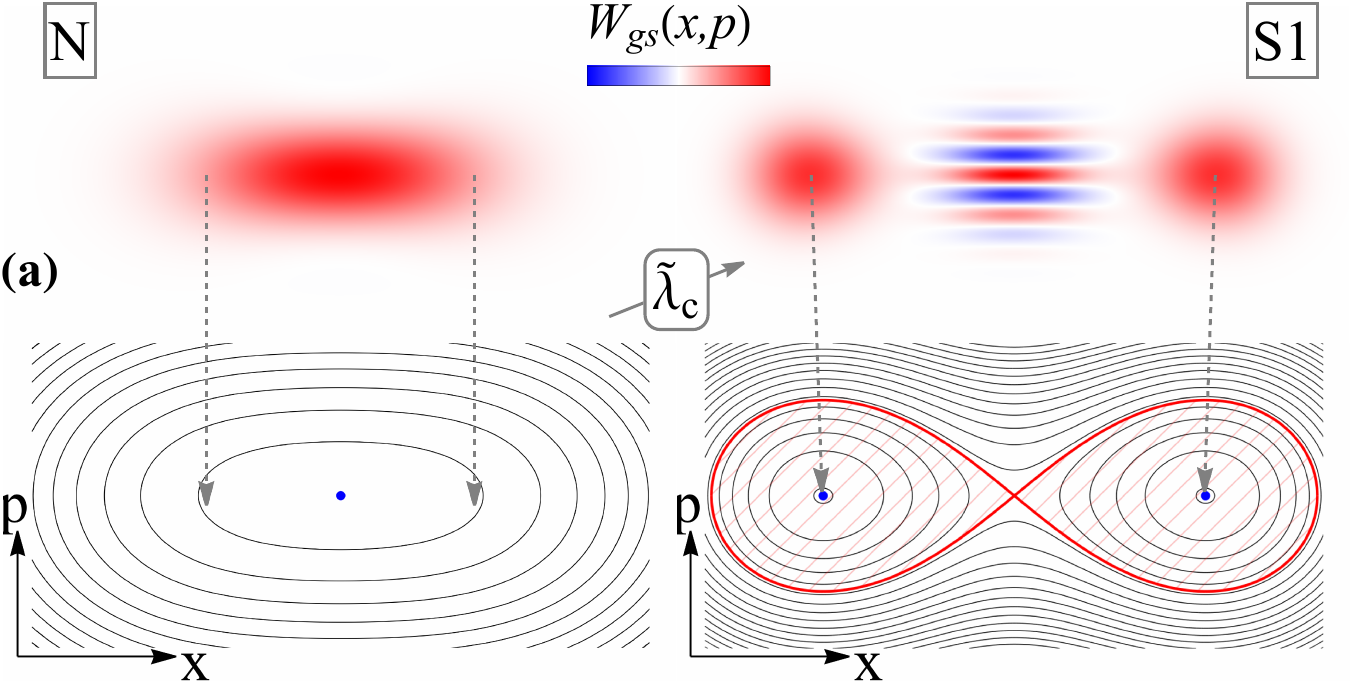}

  \vspace{0.5cm}

  \includegraphics[width=\linewidth]{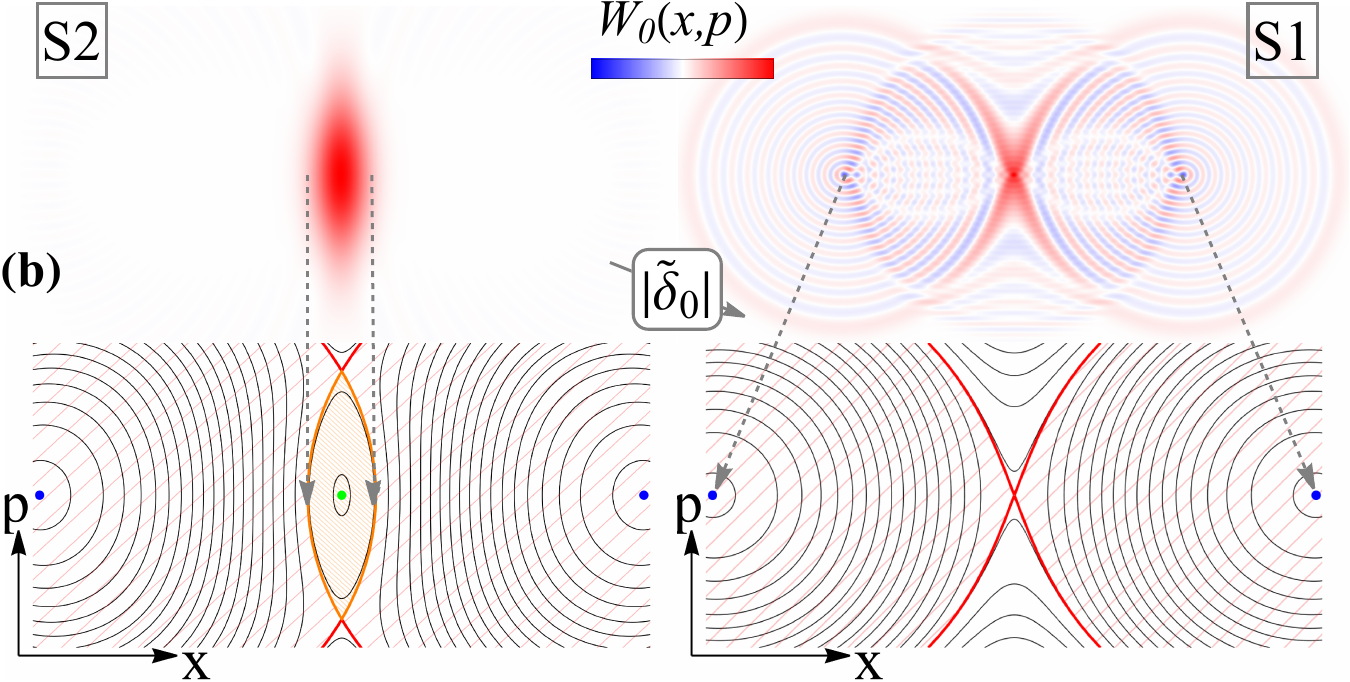}
  
  \label{fig:1}
\caption{
Critical processes in the extended Rabi model (ERM): The \textbf{N}$\to$\textbf{S1} transition (ground state QPT), panel \textbf{(a)}, occurs when the single minimum (blue point) of the semiclassical Hamiltonian $h_\downarrow(x,p;\bm{\lambda})$~\eqref{77} splits into two, leaving behind a separatrix (red curve), which we highlight by the contours corresponding to classical orbits. In accordance, the ground-state Wigner function $W_{gs}$ transforms from $p$-squeezed vacuum (left) to a cat state (right). In the \textbf{S2} phase, see panel \textbf{(b)} left, the $x$-squeezed vacuum state is stabilized~\cite{Stransky21,Kuchar26} by the local maximum (green point), energetically above the inner segment of the separatrix (orange curve). The \textbf{S2}$\to$\textbf{S1} transition (ESQPT) brings the two saddle points together, causing the stable region around the local maximum $x=p=0$ to disappear and the state with maximal vacuum contribution consequently spreads outward, as displayed by $W_{0}$ in \textbf{(b)}, right.}
\end{figure}

The standard procedure for gaining information on the ground-state of \eqref{1}, in the $\Delta\gg1$ limit, near the \textbf{N}$\to$\textbf{S1} boundary, is to perform the Schrieffer-Wolff (S-W) transformation to eliminate off-diagonal terms \cite{Hwang15,Gietka26} and obtain the new Hamiltonian $\hat{h}'$. In leading order $O(\langle\hat{n}\rangle/\Delta)$, the qubit-down projected Hamiltonian can be diagonalized by the Boguliubov transform $\hat{S}(\xi)$ and takes the form:
\begin{equation}\label{2}
\begin{aligned}
    &\hat{S}(\xi)\hat{h}_\downarrow'\hat{S}^\dagger(\xi)=\frac{1}{2}\left(\frac{\lambda^2\delta-1}{\Delta}-1\right)+\Omega\left(\frac{1}{2}+\hat{a}^\dagger\hat{a}\right)\,,\\
    &\Omega=\frac{\rm sign(1-\lambda^2\delta^2)}{\Delta}\sqrt{(\lambda^2\delta^2-1)(\lambda^2-1)}\,,\\
    &\xi=\frac{1}{4}\ln\left(\frac{\lambda^2\delta^2-1}{\lambda^2-1}\right)\,.
\end{aligned}    
\end{equation}
\noindent 
The real squeezing parameter $\xi$ entering \mbox{$\hat{S}(\xi)=\exp\left[\frac{\xi}{2}\left(\hat{a}^2-\hat{a}^{\dagger\,2}\right)\right]$} and the effective frequency $\Omega$ are well-known \cite{Hwang15,Mihailescu26} in the special case of the standard Rabi model with $\delta=0$. If we exclude $|\delta|=1$ [$U(1)$ symmetrical (anti-)Jaynes-Cummings limits of \eqref{1}], then in \textbf{N}, \eqref{2} describes a $p$-squeezed oscillator ($\xi>0$) with vanishing frequency when $\lambda\to1^-$, resulting in an infinitely squeezed vacuum as the ground-state. In the \textbf{S1} phase, the S-W transform must be preceded by a displacement [see Fig. 1(a)], since the bosonic population in the ground state is $O(\Delta)$.

In this paper, we leverage the criticality-induced squeezing at the \textbf{S2}$\rightarrow$\textbf{S1} transition, corresponding to an ESQPT, where the effective negative-frequency solution ($\langle\hat{a}^\dagger\hat{a}\rangle=o(\Delta)$) contained in \eqref{2} is of particular interest. The $\xi$-squeezed bosonic vacuum state represents here the 'anti-ground' state of \textbf{S2} and $\xi\to-\infty$ at the \textbf{S2}$\to$\textbf{S1} boundary, leading to infinite squeezing along the $x$-direction. If we moved through this boundary into \textbf{S1}, these stabilized \cite{Stransky21} squeezed states would cease to exist, and the states with maximal vacuum contribution will be spread across the available phase space [see Fig. 1(b)]. The dynamical influence of the \textbf{S2} eigenstates of \eqref{1} enclosed within the separatrix around the local maximum (see Fig. 1(b)) was recently studied \cite{Kuchar26}, but their significance in quantum metrology has not been explored.

{\it Connecting QFI with QPTs
—}Although there exist various equivalent expressions for the QFI that a pure state carries with respect to an estimated parameter, we will later utilize the connection with the quantum geometric tensor (QGT) \cite{paris2009quantum,Bengtsson2017,Strelecek25}. In a parameter space spanned by $\{\lambda^\mu\}$, the distance between two states separated by infinitesimal parameter variations takes the form:
\begin{equation*}
 ds^2=1-|\langle\psi(\bm{\lambda})|\psi(\bm{\lambda}+d\bm{\lambda})\rangle|^2=Q_{\mu\nu}d\lambda^\mu d\lambda^\nu \,.\end{equation*}
\noindent The QGT $Q_{\mu\nu}$ defined by the summation over individual variations can be written as
\begin{equation*}
    Q_{\mu\nu}=\langle\partial_\mu\psi(\bm{\lambda})|(1-|\psi\rangle\langle\psi|)|\partial_\nu\psi(\bm{\lambda})\rangle\,.
\end{equation*}
By comparison \cite{paris2009quantum}, we see that the QFI carried by $|\psi\rangle$ in the process of estimating $\lambda^\mu$ is simply $\mathcal{I}_\mu(\psi)=4Q_{\mu\mu}$. From direct evaluation of the QGT for the \eqref{2} solution, we obtain the following single-parameter QFIs:
\begin{equation}\label{5}
\begin{aligned}
\mathcal{I}_\lambda(\psi_n)/I_n&=4\left(\frac{\partial\xi}{\partial\lambda} \right)^2=\frac{\lambda^2(1-\delta^2)^2}{(\lambda^2-1)^2(\lambda^2\delta^2-1)^2}\,,\\
\mathcal{I}_\delta(\psi_n)/I_n&=4\left(\frac{\partial\xi}{\partial\delta} \right)^2=\frac{\lambda^4\delta^2}{(\lambda^2\delta^2-1)^2}\,,
\end{aligned}
\end{equation}
\noindent where $|\psi_n\rangle=\hat{S}^\dagger(\xi)|n\rangle$, $I_n=(n^2+n+1)/2$ and \mbox{$n\in \mathbb{N}_0$ correspond to squeezed Fock states.} 
Notice that for $(\lambda, \delta) \in \mathbf{N} \cup  \mathbf{S2}$, Heisenberg scaling $\mathcal{I}_\mu\sim N_\mu^2T^2$ holds near critical points, where the time scale $T\equiv(\Omega/\Delta)^{-1}$ and \mbox{$N_\mu\equiv\langle\alpha_\mu\hat{x}^2+\beta_\mu\hat{p}^2\rangle$} is an effective resource for suitable $\alpha_\mu,\,\beta_\mu$~\cite{Mihailescu26, Gietka26}.

The key aspect of expressions \eqref{5} is that choosing $(\lambda_N,\delta_N)\in\;$\textbf{N} and $(\lambda_2,\delta_2)\in\;$\textbf{S2} such that \mbox{$|\lambda_N-1|=\varepsilon=|\lambda_2|\delta_2|-1|$,} where $\varepsilon\ll1$, we have \mbox{$|\partial\xi/\partial\lambda|_N\approx1/(4\varepsilon)\approx|\partial\xi/\partial\delta|_2$}. The identical scaling of $\mathcal{I}_\lambda$ near the \textbf{N}$\to$\textbf{S1} boundary and $\mathcal{I}_\delta$ near the \textbf{S2}$\to$\textbf{S1} is utilized below as a way to compare state-preparation protocols in terms of criticality-enhanced sensing capabilities. Furthermore, we can use the solution \eqref{2} to link \eqref{5} to a class of observables $\hat{O}$ satisfying the maximal signal-to-noise ratio (SNR) condition \cite{braunstein1994statistical,Garbe20,Mihailescu26}:
\begin{equation}\label{6}
    \mathcal{I}_\mu(\psi(\bm{\lambda}))=\max_{\hat{O}} \left[ \text{SNR}_\mu(\bm{\lambda};\hat{O}) \right]\equiv\max_{\hat{O}} \left[  \frac{\left|\partial_\mu\langle\hat{O}\rangle_{\psi(\bm{\lambda})}  \right|^2}{\text{Var}_{\psi(\bm{\lambda})}(\hat{O})} \right].
\end{equation}
\noindent By direct evaluation, we find that for $n\in\{0,1\}$, the QFIs \eqref{5} are reached through \eqref{6} by the entire class of observables $\hat{O}=\alpha \hat{x}^2+\beta\hat{p}^2$, where $\alpha,\,\beta\in\mathbb{R}$. Therefore, we already know the optimal measurement strategies to leverage the criticality enhanced precision of the ground and 'anti-ground' states in \textbf{N} and \textbf{S2}, respectively.

\begin{figure}[t!]
  \centering
  \includegraphics[width=\linewidth]{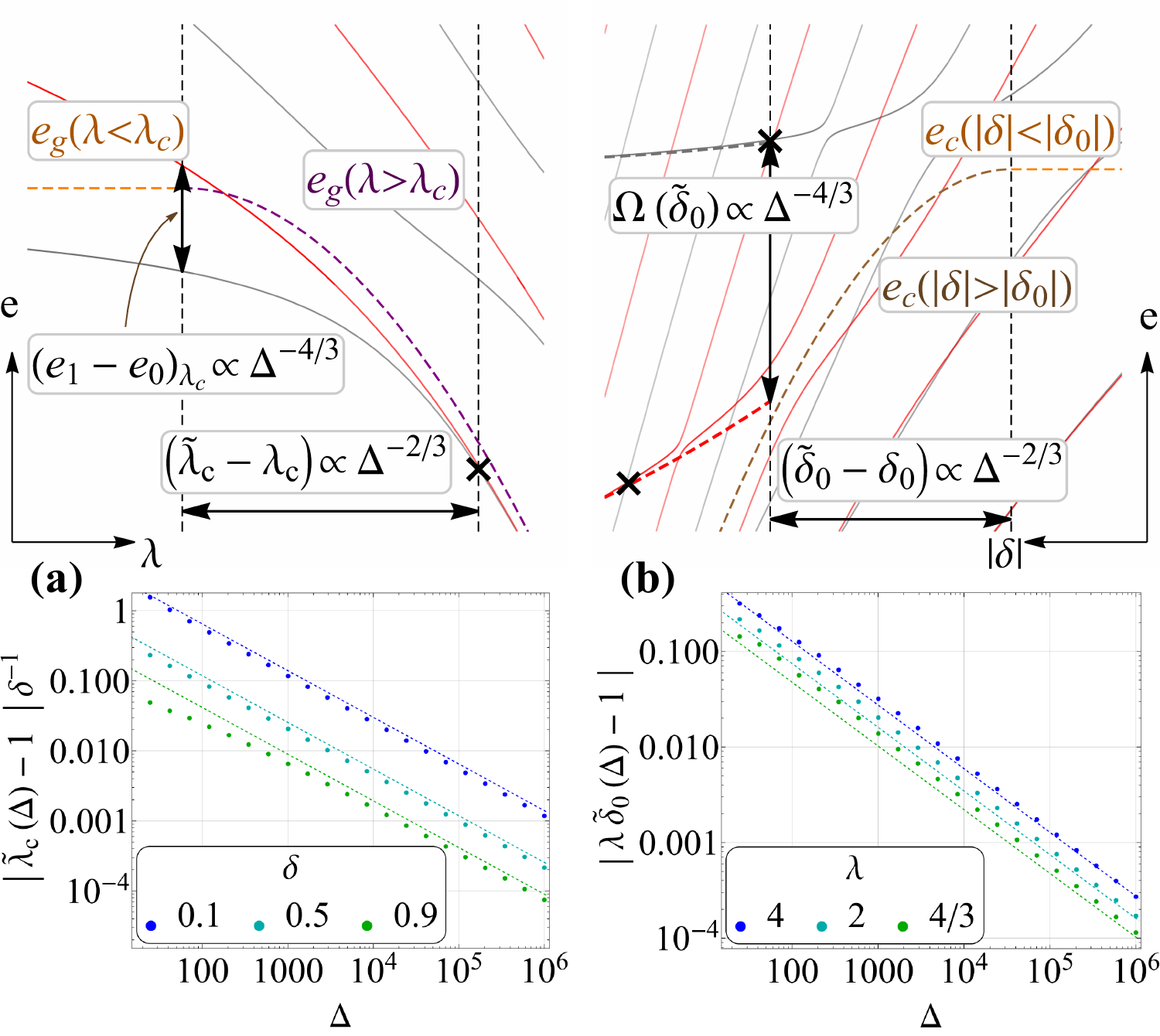}
  \label{fig:2}
  \caption{Finite-$\Delta$ scaling of critical point positions (horizontal arrows) and energy gaps (vertical arrows): In \textbf{(a)}, top, the spectral $\lambda$-dynamics shows that the \textbf{N}$\to$\textbf{S1} corrected boundary $\tilde{\lambda}_c$ \eqref{7} moves beyond $\lambda_c$ (black cross placed at the merger of opposite parity energy levels---gray for positive and red for negative parity), signifying the end of criticality-induced squeezing. At $\tilde{\lambda}_c$, the energy gap closes and the ground-state reaches the energy of the semiclassical minimum (dashed curves: orange in \textbf{N} and purple in \textbf{S1} phases), as $\Delta\rightarrow\infty$. In \textbf{(b)}, top, the corrected \textbf{S2}$\to$\textbf{S1} boundary, marking the disappearance of the stabilized squeezed vacuum (black cross on gray-dashed line), satisfies $|\tilde{\delta}_0|>|\delta_0|$ and while the energy gap at $\tilde{\delta}_0$ closes with increasing $\Delta$, the residual stabilized squeezed state is destroyed prior to reaching it (black cross on red-dashed line). These processes are bound from below by the semiclassical saddle-point energy (brown- and orange-dashed lines). 
  Bottom panels verify the~\eqref{7} approximation (dashed lines) against numerically obtained coordinates of the maximal ground-state squeezing (points) for both \textbf{N}$\to$\textbf{S1} \textbf{(a)} and \textbf{S2}$\to$\textbf{S1} \textbf{(b)}.}
\end{figure}

{\it Finite-$\Delta$ ground- and excited-state QPT processes—}To guide the design of feasible metrological protocols that utilize near-boundary critical states, it is crucial to evaluate finite-size corrected critical points $\tilde{\lambda}_c(\Delta)$ \mbox{($\lambda_c=1$)} and $\tilde{\delta}_0(\Delta)$ ($|\delta_0|=\lambda^{-1}$). A practical approach is to examine the available phase-space volume and determine boundaries for existence of the squeezed-vacuum ('anti-') ground state. 
 In the course of QPT protocols, beyond the \textbf{N}$\to$\textbf{S1} boundary, $\lambda_c$, opposite parity ($\mathbb{Z}_2$ symmetry of \eqref{1} \cite{CarrBook,SachdevBook,Hwang15}) eigenvalues merge [see Fig. 2(a)], which corresponds to the splitting of the ground state into the double-well structure [see Fig. 1(a)]. The $\xi$-squeezed vacuum will therefore cease to be the ground state at $\tilde{\lambda}_c(\Delta)>\lambda_c$, if the phase-space volume of the double-well [hatched red region in Fig. 1(a)] becomes large enough to host two states. On the other hand, when approaching the ESQPT at the \textbf{S2}$\to$\textbf{S1} boundary, $\delta_0$, the phase-space region bound by the separatrix and containing the local maximum [hatched orange region in Fig. 1(b)] decreases in volume  until at $|\tilde{\delta}_0|(\Delta)>|\delta_0|$, it can no longer support the squeezed vacuum state. The eigenstate with maximal vacuum contribution then 'leaks out' of this region along the outer separatrix [see Fig. 1(b)]. In spectral dynamics, this manifests itself as the widening of the same-parity avoided crossings [see Fig. 2(b) top] due to proximity to the saddle-point energy, 
  eventually leading to the destruction of the stabilized critical excited states contained in the \eqref{2} solution.

The phase-space volumes discussed above and highlighted in Fig. 1 can be obtained analytically by integration of the energy shells defined by the semiclassical Hamiltonian \cite{Stransky21, Kuchar26}:
\begin{equation}\label{77}
    h_\downarrow(x,p;\bm{\lambda})=\frac{x^2+p^2}{2}-\frac{1}{2}\sqrt{2\lambda^2(x^2+\delta^2p^2)+1}\:.
\end{equation}
\noindent The red and orange-hatched regions in Fig. 1 yield phase-space volumes $v_{N\to S1}(\bm{\lambda}_1)$ and $v_{S2\to S1}(\bm{\lambda}_2)$, respectively, where $\bm{\lambda}_1\in\,$\textbf{S1} and $\bm{\lambda}_2\in\,$\textbf{S2}. By inverting the conditions \mbox{$v_{N\to S1}(\tilde{\lambda}_c,\delta)=\frac{4\pi}{\Delta}\,,$} \mbox{$v_{S2\to S1}(\lambda,\tilde{\delta}_0)=\frac{2\pi}{\Delta}\,,$} we arrive at the finite-size corrected critical points:

\begin{equation}\label{7}
    \begin{aligned}
    \tilde{\lambda}_c(\Delta;\delta)&=1+\left(\frac{3\pi}{4\Delta}\sqrt{\frac{1-\delta^2}{2}}\right)^{2/3}+O\left(\Delta^{-4/3}\right)\,,\\
    |\tilde{\delta}_0|(\Delta;\lambda)&=\lambda^{-1}+\left(\frac{3\pi}{8\Delta}\sqrt{\frac{\lambda^2-1}{\lambda^3}}\right)^{2/3}+O\left(\Delta^{-4/3}\right)\,.
    \end{aligned}
\end{equation}
\noindent Through numerical diagonalization of \eqref{1} in the vicinity of infinite-size critical points, $\lambda_c$ and $\delta_0$, we probe the parametric coordinates~\eqref{7} corresponding to the local maxima of ratios $\langle\hat{x}^2\rangle/\langle\hat{p}^2\rangle$ and $\langle\hat{p}^2\rangle/\langle\hat{x}^2\rangle$, respectively. These maxima mark the end of criticality-induced squeezing at given $\Delta$ and from Fig. 2 (bottom), we see that their coordinates $\tilde{\lambda}^\mu$ are well predicted by \eqref{7} in both the \textbf{N}$\rightarrow$\textbf{S1} (left) and \textbf{S2}$\rightarrow$\textbf{S1} (right) cases.

Although in both cases, 
 the finite-size corrections in \eqref{7} behave as $O(\Delta^{-2/3})$ and also 
the key energy gaps (see Fig. 2) scale as $O(\Delta^{-4/3})$ \cite{Hwang15}, the underlying mechanisms relevant to critical-state synthesis differ significantly: Upon approaching the \textbf{N}$\to$\textbf{S1} boundary, the effective oscillator frequency $\Omega$ approaches zero \eqref{2}, but apart from $\langle\hat{a}^\dagger\hat{a}\rangle\ll\Delta$, there is no spectral restriction on the number of states experiencing criticality-induced squeezing. Therefore, for $\Delta\gg1$, the solution \eqref{2} approximates the low-$\langle\hat{n}\rangle$ eigenstates of \eqref{1} in \textbf{N}, and these states remain coupled near the boundary, obstructing the desired adiabatic evolution. In contrast, for \textbf{S2} and $\Delta\gg1$, the part of the spectrum approximated by \eqref{2} is bound from below by the semiclassical saddle point energy \cite{Stransky21,Kuchar26} $e_c(\bm{\lambda})=-\left[(\lambda\delta)^2+(\lambda\delta)^{-2})\right]/4$ and from above by the local maximum. Crucially, since the fundamental limitation placed on the number of critically squeezed excited states is the extent of the phase-space region described by $v_{S2\to S1}(\bm{\lambda})$, there cannot be any states approximated by \eqref{2} outside this region. For finite $\Delta$, there exists 
a $\tilde{\lambda}^\mu$-neighborhood of the finite-size \textbf{S2}$\to$\textbf{S1} boundary, where only one such state can exist (see Fig. 2(b)). Therefore, in this neighborhood, the state does not experience the conventional slowing-down of the dynamics, because the diabatic passage through avoided crossings leads to criticality-induced squeezing, while any coupling to background states is weak due to significant structural differences. We explore this behavior in more detail below.

\begin{figure}[t!]
  \centering
  \includegraphics[width=\linewidth]{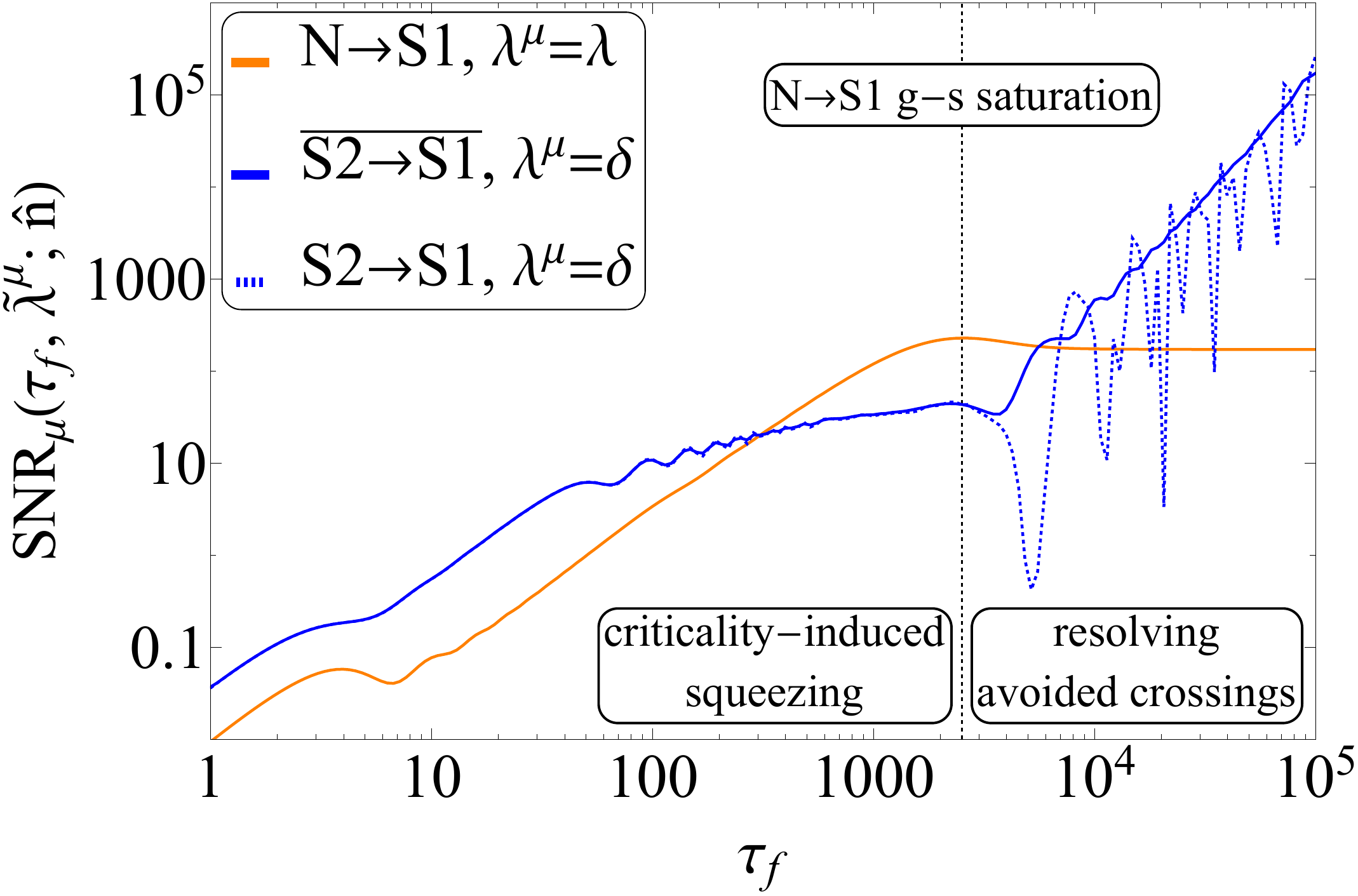}
  \caption{Signal-to-noise ratios (SNRs) of \eqref{8} for linear-in-time parametric ramps, 
  as functions of the ramp duration $\tau_f$. All protocols start from the qubit-down and motional vacuum state $|\psi_i\rangle=|{\small\downarrow}\,,0\rangle$ and the system size is set to $\Delta=25$. The \textbf{N}$\to$\textbf{S1} protocols proceed at fixed $\delta=0$, start at $\lambda_i=0$ and end at \mbox{$\lambda_f=\tilde{\lambda}_c\doteq1.165$}. The corresponding $\text{SNR}_\lambda$ reaches its maximum at \mbox{$\tilde{\tau}_f\approx10^{3.4}$,} (black-dashed vertical line). The \textbf{S2}$\to$\textbf{S1} protocols proceed at fixed $\lambda=2$, start at $\delta_i=1$ and end at \mbox{$\delta_f=\tilde{\delta}_0\doteq0.601\,$.} While the ground-state protocols amount to gradually increasing fidelity of the near-boundary squeezed state [cf. Fig.1(a), left], the ESQPT protocols at durations $\tau_f\gtrsim\tilde{\tau}_f$ start to resolve subtle spectral features [cf. Fig.2(b), top] and the $\text{SNR}_\delta$ grows rapidly with $\tau_f$. The solid blue line averages the $\text{SNR}_\delta$ dependencies by uniformly sampling the end points \mbox{$(\delta_f-\tilde{\delta}_0)\sim\mathcal{U}(-d\delta,d\delta)$},  $d\delta=10^{-2}$.}
  \label{fig:1}
\end{figure}

{\it Comparison of accessible resources—}To compare the metrological utility of the processes described above, we evaluate
the dependencies of signal-to-noise ratios for the phonon number operator $\hat{n}$
\begin{equation}\label{8}
\text{SNR}_\mu(\tau_f,\bm{\lambda}_f;\hat{n})=\frac{\left|\partial_\mu\langle\hat{n}\rangle_{\psi(\tau_f,\bm{\lambda}_f)}\right|^2}{\text{Var}_{\psi(\tau_f,\bm{\lambda}_f)}(\hat{n})}
\end{equation}
\noindent on characteristic durations $\tau_f$ of parametric trajectories $\lambda^\mu_0\overset{\tau_f}{\longrightarrow}\lambda^\mu_f\,.$ The choice $\hat{O}=\hat{n}$ provides the simplest option which complies with the family of operators satisfying \eqref{6} for QFIs \eqref{5} and is preserved by the usual unitary transformations necessary to acquire the mapping \cite{Pedernales16,Duan21,Kuchar26} of a realistic qubit-phonon system to the ERM model Hamiltonian \eqref{1}. For a general time-dependent trajectory $\bm{\lambda}(\tau,\bm{\lambda}_f)$ with a fixed end-point $\bm{\lambda}_f$, determining \eqref{8} requires to solve the equations
\begin{equation}\label{9}
    \begin{aligned}
\left[i\frac{\partial}{\partial\tau}-\hat{h}(\bm{\lambda}(\tau,\bm{\lambda}_f))\right]|\psi(\tau,\bm{\lambda}_f)\rangle&=0\,,\;|\psi(0,\bm{\lambda}_f)\rangle\equiv|\psi_i\rangle\,,\\
\left[i\frac{\partial}{\partial\tau}-\hat{h}(\bm{\lambda}(\tau,\bm{\lambda}_f))\right]|\chi_\mu(\tau,\bm{\lambda}_f)\rangle&=\hat{d}_\mu(\tau,\bm{\lambda}_f)|\psi(\tau,\bm{\lambda}_f)\rangle\,,
\end{aligned}
\end{equation}
\noindent where we denoted $\hat{d}_\mu(\tau,\bm{\lambda}_f)=\partial\hat{h}(\bm{\lambda}(\tau,\bm{\lambda}_f))/\partial\lambda^\mu_f$, \mbox{$|\chi_\mu(\tau,\bm{\lambda}_f)\rangle=\partial|\psi(\tau,\bm{\lambda}_f)\rangle/\partial\lambda^\mu_f$} and $|\chi_\mu(0,\bm{\lambda}_f)\rangle=0\:.$ To accommodate experimental capabilities \cite{Pedernales16,Duan21,Leibfried2003} and to do an illustrative comparison of the \textbf{N}$\to$\textbf{S1} and \textbf{S2}$\to$\textbf{S1} protocols, we chose the initial state $|\psi_i\rangle=|{\small\downarrow}\,,0\rangle$ and parametric trajectories linear in time \mbox{$\lambda^\mu(\tau,\bm{\lambda}_f)=\lambda^\mu_i+(\lambda^\mu_f-\lambda^\mu_i)\,\tau/\tau_f\,.$}  The match~\eqref{5} in scalings of $\mathcal{I}_\lambda$ and $\mathcal{I}_\delta$ with distance from the boundaries \textbf{N}$\to$\textbf{S1} and \textbf{S2}$\to$\textbf{S1}, respectively, naturally motivates the choice $\lambda^\mu=\lambda$ for the first protocol and $\lambda^\mu=\delta$ for the latter. Furthermore, assuming that $\lambda_i=0$ for \textbf{N}$\to$\textbf{S1} and $\delta_i=1$ for \textbf{S2}$\to$\textbf{S1}, $\psi_i$ will be an initial eigenstate of \eqref{1} in both protocols with zero initial squeezing 
$\xi(\bm{\lambda}_i)=0 $ in both cases.

In Fig. 3, we see the quantitative comparison of metrological resources provided by the driving protocols as functions of their durations. We used the same system size $\Delta=25$ for both criticality-induced processes, the {color{red}(optimal)} fixed $\delta=0$ for the \textbf{N}$\to$\textbf{S1} squeezing and a moderate \cite{Duan21,Kuchar26} fixed $\lambda=2$ for \textbf{S2}$\to$\textbf{S1}. The protocol end-points $\lambda^\mu_f$ were taken as the $O(\Delta^{-4/3})$ corrected versions of \eqref{7}. The results in Fig. 3 show that depending on $\tau_f$, both processes can be separated into two regimes, see Fig. 3, by durations $\tilde{\tau}_f^\mu$, which for our choice of parameters approximately merge into a single $\tilde{\tau}_f$ represented by the dashed vertical line. For both ground- and excited-state transitions, criticality-induced squeezing takes place at $\tau_f\lesssim\tilde{\tau}_f^\mu$, beyond which the \textbf{N}$\to$\textbf{S1} SNR saturates by reaching the ground state at $\lambda_f$ with high fidelity. Although $\text{SNR}_\delta$ in the \textbf{S2}$\to$\textbf{S1} protocol initially grows faster than $\text{SNR}_\lambda$ in \textbf{N}$\to$\textbf{S1}, it appears to saturate below the ground state value. We note that the final achievable squeezing parameter in the ESQPT protocol also scales with $\lambda\,$:
\begin{equation}\label{10}
|\xi(\lambda,\tilde{\delta}_0(\Delta))|\approx\ln\left[\frac{2\sqrt2}{3\pi}\Delta(\lambda^2-1)\right]/6\,.
\end{equation}
\noindent However, before the true saturation of $\text{SNR}_\delta$ occurs, it starts to significantly oscillate and increase with $\tau_f>\tilde{\tau}_f^\delta$, 
which we emphasize by including $\overline{\text{SNR}}_\delta$ averaged over $\delta$ in the vicinity of $\delta_f$. It is apparent that the averaging does not impact the criticality-induced squeezing, yet it smooths out the oscillations of the long-$\tau_f$ $\text{SNR}_\delta$, while preserving the steep increase with the duration $\tau_f$. This behavior is attributed to the ability of $|\chi_\delta(\tau)\rangle$, see~\eqref{9}, to resolve individual narrow avoided crossings when passing through them sufficiently 
slowly: The widths and positions of the avoided crossings induce 
a substantial $\delta$-sensitivity imprinted by the saddle-point type ESQPT (cf. Figs. 1 and 2). The wavefunction also starts to partially cross the avoided crossings adiabatically, causing it to 'leak' into the background states (Fig. 1(b), right). These combined effects lead to sensitivity significantly exceeding what can be achieved by the criticality-induced squeezing alone. 

\begin{figure}[t!]
  \centering
  \includegraphics[width=\linewidth]{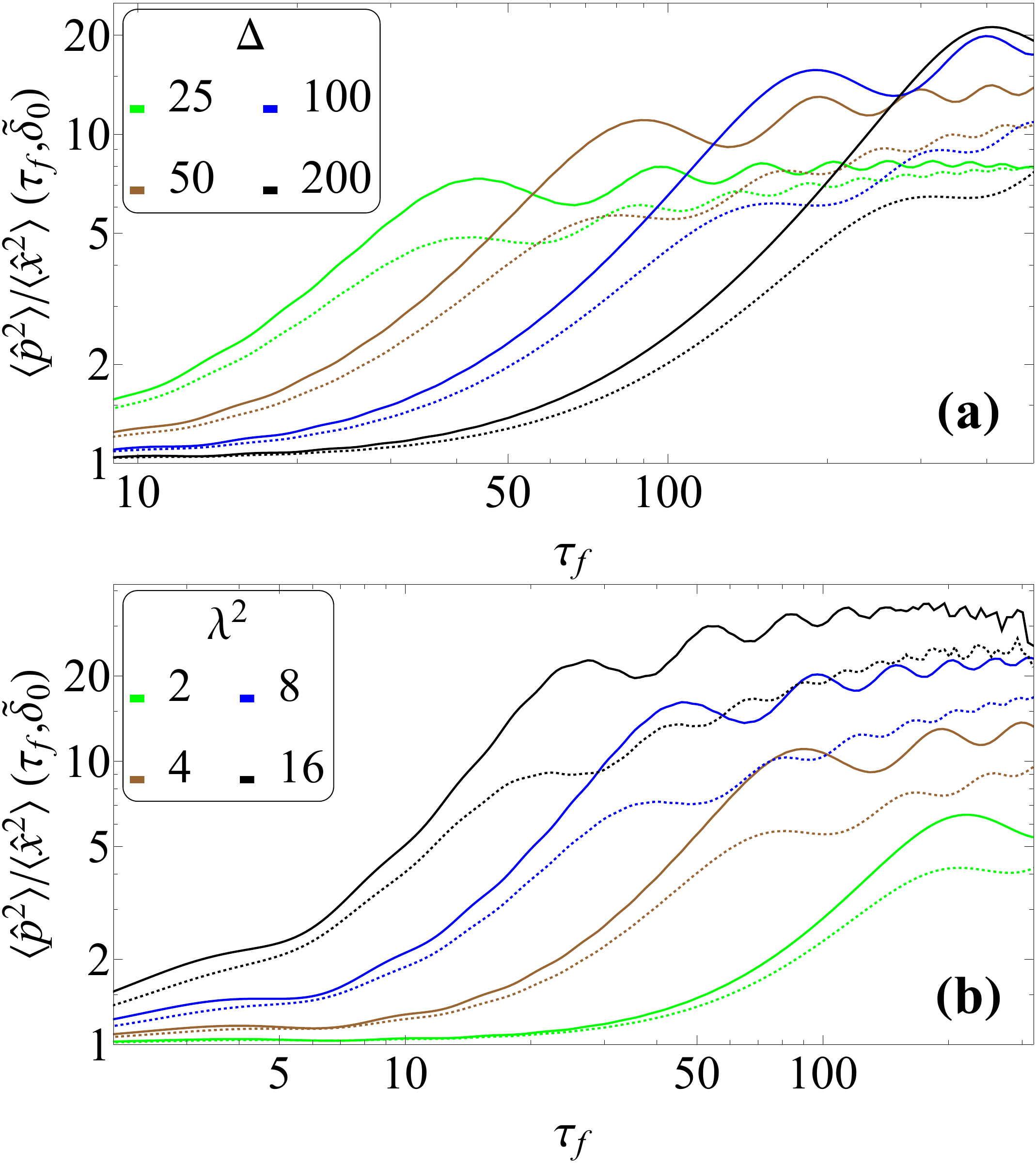}
  \caption{Comparison of linear-in-time (dashed lines) and geodesic (solid lines) trajectories in the \textbf{S2}$\to$\textbf{S1} process of squeezing the initial qubit-down vacuum prepared at \mbox{$\delta_i=1$}: We plot the quadrature ratio $\langle\hat{p}^2\rangle/\langle\hat{x}^2\rangle$ at the end of the protocol lasting $\tau_f$ and reaching $\delta_f=\tilde{\delta}_0(\Delta;\lambda)$ (see \eqref{7}), at fixed $\lambda=2$ (panel a) or $\Delta=50$ (b).}
  \label{fig:1}
\end{figure}

Finally, we elaborate on the acceleration of \textbf{S2}$\to$\textbf{S1} squeezing  considering geodesic driving~\cite{Bengtsson2017,Strelecek25}. Since residual critical excited states vanish in close proximity to the boundary, optimal state preparation should spend an extended period of time in the $\tilde{\lambda}^\mu$-neighborhood, where $v_{S2\to S1}(\bm{\lambda})$ supports only a single quantum state. Due to the finite-size corrected critical point moving deeper into \textbf{S2} (contrary to what happens in \textbf{N}$\to$\textbf{S1}), an optimized parametric trajectory can be obtained naturally from the solution \eqref{2}. The QGT determined by \eqref{2} for the squeezed vacuum can be diagonalized and has a single non-trivial eigenvalue: \mbox{$ds^2=\frac{1}{2}d\xi^2\,.$} Traversing the shortest possible distance on the squeezed vacuum manifold in finite time $\tau_f$ therefore leads to geodesics \mbox{$\xi[\lambda(\tau),\delta(\tau)]=\xi_i+(\xi_f-\xi_i)\tau/\tau_f$.} To provide a tangible comparison with linear-in-time protocols in \mbox{Fig. 3,} we set $\lambda(\tau)\equiv\lambda$, $\delta(0)=1$, which leads to the following geodesic:
\begin{equation}\label{11}
    \delta_g(\tau;\bm{\lambda}_f)=\sqrt{\lambda^{-2}+(\delta_f^2-\lambda^{-2})^{\tau/\tau_f}(1-\lambda^{-2})^{1-\tau/\tau_f}}\:.
\end{equation}
\noindent We note here that assuming a parametric trajectory highly sensitive to variations of $\bm{\lambda}_f$, such as a near-critical geodesic [$|\delta_f|\approx\lambda^{-1}$ in \eqref{11}], further accelerates the growth of $||\chi_\mu||$ through \eqref{9}. This could enhance the sensitivity in finite-duration driving protocols accompanied by criticality even more, provided that such parametric trajectories can be physically conditioned to depend on the estimated parameter (end-point position).

In Fig. 4, we compare criticality-induced squeezing in \textbf{S2}$\to$\textbf{S1} protocols, obtained by linear-in-time parametric trajectories and fixed-$\lambda$ geodesics described in \eqref{11}. We again chose $\delta_f=\tilde{\delta}_0$ and examined scaling with system size $\Delta$ and coupling strength $\lambda$. The quantity of interest is the ratio of quadratures $\langle\hat{p}^2\rangle/\langle\hat{x}^2\rangle$, which yields $\exp(-4\xi)$ for the eigenstates described by \eqref{2}. In both Figs. 4(a) and (b), we show an enhancement by driving along geodesics compared to linear-in-time protocols in the squeezing regime. Observing the $\lambda=2$ geodesic drivings in Fig. 4(a) and the change in $\tau_{\rm max}(\Delta)$, corresponding to the first local maxima after the initial steep growth in the quadrature ratio, we find \mbox{$\tau_{\rm max}\sim\Delta^{1.04(3)}$} and \mbox{$\left(\langle\hat{p}^2\rangle/\langle\hat{x}^2\rangle\right)_{\rm max}\sim\Delta^{0.55(4)}$} in the range $\Delta\in(25,400)$. Although the final saturation value of the latter should scale as $\Delta^{2/3}$, predicted by \eqref{10}, the reduction in temporal scaling due to geodesics is significant. For comparison, in the linear-in-time \textbf{N}$\to$\textbf{S1} protocols with $\delta=0$, $\lambda_i=0$, $\lambda_f=\tilde{\lambda}_c$ and identical $\Delta$-range, the saturation duration exhibits \mbox{$\tilde{\tau}_f^\lambda\sim\Delta^{1.98(1)}$} and the inverted quadrature ratio \mbox{$\left(\langle\hat{x}^2\rangle/\langle\hat{p}^2\rangle\right)_{\rm sat}\sim\Delta^{0.62(1)}$.}

Lastly, in Fig. 4(b), we see that generation of highly-squeezed states in the \textbf{S2}$\to$\textbf{S1} protocols does not need to be paid for in system size $\Delta$ and preparation time $\tau_f$, but can be achieved by increasing the coupling strength $\lambda$, which brings notable advantages independent of the driving scheme. Not only does it in turn increase the achievable squeezing through \eqref{10}, but the required duration also reduces as the energy gaps widen with increasing $\lambda$. However, this widening also impacts the avoided crossings near finite-size corrected critical points, which causes the crossover to adiabatic passage to occur at shorter $\tilde{\tau}_f^\delta$ as well (see $\lambda=4$ in Fig. 2(b)). We emphasize that similar enhancement by increased coupling strength $\lambda$ is not possible for ground-state QPT \textbf{N}$\to$\textbf{S1} protocols, as the only free interaction parameter here is $\delta$, and maximum squeezing at given $\Delta$ is attained for $\delta=0$.

{\it Conclusion—}Our finite-size analysis and numerical results establish a direct comparison of ground- and excited-state QPTs as metrological resources in terms of protocol-duration-dependent signal-to-noise ratios for interaction-parameter estimation using the experimentally accessible mean bosonic population \cite{Pedernales16,Duan21,Stransky21,Kuchar26,Leibfried2003}. We identified a long-time regime in which ESQPT-based sensing surpasses ground-state sensitivity by combining diabatic and adiabatic passage through narrow avoided crossings near the critical energy. Furthermore, we showed that approaching the ESQPT increasingly closely with growing system size imprints progressively stronger squeezing onto the driven state, at the cost of longer preparation times. However, this time scaling can be softened through analytically derived geodesics or even reversed by instead increasing the overall coupling strength, offering a way to circumvent the critical slowing down that limits conventional critical quantum metrology \cite{Frerot18,Garbe20,Mihailescu26,Gietka26}.

{\it Acknowledgments—}We thank P. Cejnar, P. Str\'ansk\'y, L. Slodi{\v c}ka, O. {\v C}\'ip, O. {\v C}ernot\'ik, R. Filip, V. {\v S}varc and A. Rela\~no for insightful discussions and comments. This work was supported by the Czech Ministry of Education, Youth and Sports, project No. CZ.02.01.01/00/22 008/0004649 (QUEENTEC).
Authors also acknowledge the project 23FUN03 HIOC, which has received funding from the European Partnership on Metrology, co-financed from the European Union’s Horizon Europe Research and Innovation Programme and by the Participating States.
M.K. acknowledges partial support by the Czech Science Foundation project No. 25-16056S, and travel funds by the Czech Academy of Sciences via the ``Akademie budoucnosti'' program.


\begin{thebibliography}{99}

\bibitem{braunstein1994statistical}
S. L. Braunstein and C. M. Caves,
Phys. Rev. Lett. \textbf{72}, 3439 (1994).

\bibitem{giovannetti2004quantum}
V. Giovannetti, S. Lloyd, and L. Maccone,
Science \textbf{306}, 1330 (2004).

\bibitem{paris2009quantum}
M. G. A. Paris,
Int. J. Quantum Inf. \textbf{7}, 125 (2009).

\bibitem{CarrBook}
L. D. Carr, ed.,
\textit{Understanding Quantum Phase Transitions}
(CRC Press, Boca Raton, 2010).

\bibitem{SachdevBook}
S. Sachdev,
\textit{Quantum Phase Transitions}, 2nd ed.
(Cambridge University Press, Cambridge, 2011).

\bibitem{Frerot18}
I. Fr\'erot and T. Roscilde,
Phys. Rev. Lett. \textbf{121}, 020402 (2018).

\bibitem{Wang14}
T.-L. Wang et al.,
New J. Phys. \textbf{16}, 063039 (2014).

\bibitem{Wei18}
B.-B. Wei and X.-C. Lv,
Phys. Rev. A \textbf{97}, 013845 (2018).

\bibitem{Hwang15}
M.-J. Hwang, R. Puebla, and M. B. Plenio,
Phys. Rev. Lett. \textbf{115}, 180404 (2015).

\bibitem{Pedernales16}
J. S. Pedernales et al.,
Sci. Rep. \textbf{5}, 15472 (2015).

\bibitem{Duan21}
M.-L. Cai et al.,
Nat. Commun. \textbf{12}, 1126 (2021).

\bibitem{Garbe20}
L. Garbe et al.,
Phys. Rev. Lett. \textbf{124}, 120504 (2020).

\bibitem{Mihailescu26}
G. Mihailescu et al.,
PRX Quantum \textbf{7}, 020201 (2026).

\bibitem{Gietka26}
G. Mihailescu and K. Gietka,
arXiv:2602.03675 (2026).

\bibitem{Liu21}
R. Liu et al.,
npj Quantum Inf. \textbf{7}, 170 (2021).

\bibitem{Chu21}
Y. Chu et al.,
Phys. Rev. Lett. \textbf{126}, 010502 (2021).

\bibitem{Cejnar21}
P. Cejnar, P. Str\'ansk\'y, M. Macek, and M. Kloc,
J. Phys. A: Math. Theor. \textbf{54}, 133001 (2021).

\bibitem{Monodromy}
P. Cejnar, M. Macek, S. Heinze, J. Jolie, and J. Dobe\v{s},
J. Phys. A: Math. Gen. \textbf{39}, L515 (2006).

\bibitem{Caprio8}
M. A. Caprio, P. Cejnar, and F. Iachello,
Ann. Phys. (N.Y.) \textbf{323}, 1106 (2008).

\bibitem{Corps25}
\'A. L. Corps, A. Rela\~no, and J. C. Halimeh,
Phys. Rev. Research \textbf{6}, 043080 (2024).

\bibitem{Stransky14}
P. Str\'ansk\'y, M. Macek, and P. Cejnar,
Ann. Phys. (N.Y.) \textbf{345}, 73 (2014).

\bibitem{Stransky16}
P. Str\'ansk\'y and P. Cejnar,
Phys. Lett. A \textbf{380}, 2637 (2016).

\bibitem{Santos15}
L. F. Santos and F. P\'erez-Bernal,
Phys. Rev. A \textbf{92}, 050101 (2015).

\bibitem{Bernal17}
F. P\'erez-Bernal and L. F. Santos,
Fortschr. Phys. \textbf{65}, 1600035 (2017).

\bibitem{Cejnar08}
P. Cejnar and P. Str\'ansk\'y,
Phys. Rev. E \textbf{78}, 031130 (2008).

\bibitem{Zakaria21}
Z. Mzaouali et al.,
Phys. Rev. E \textbf{103}, 032145 (2021).

\bibitem{Bernal21}
Q. Wang and F. P\'erez-Bernal,
Phys. Rev. E \textbf{103}, 032109 (2021).

\bibitem{Guan21}
Q. Guan and R. J. Lewis-Swan,
Phys. Rev. Research \textbf{3}, 033199 (2021).

\bibitem{Zhou23}
L. Zhou et al.,
Phys. Rev. Research \textbf{5}, 013087 (2023).

\bibitem{Mumford26}
J. Mumford and R. J. Lewis-Swan,
Phys. Rev. A \textbf{113}, 033302 (2026).

\bibitem{Liu17}
M. Liu et al.,
Phys. Rev. Lett. \textbf{119}, 220601 (2017).

\bibitem{Stransky21}
P. Str\'ansk\'y, P. Cejnar, and R. Filip,
Phys. Rev. A \textbf{104}, 053722 (2021).

\bibitem{Kuchar26}
M. Kucha\v{r} and M. Macek,
Phys. Rev. Research \textbf{8}, 033241 (2026).

\bibitem{Bengtsson2017}
I. Bengtsson and K. {\.Z}yczkowski,
\textit{Geometry of Quantum States:
An Introduction to Quantum Entanglement},
2nd ed. (Cambridge University Press, Cambridge, 2017).

\bibitem{Strelecek25}
J. St\v{r}ele\v{c}ek and P. Cejnar,
Phys. Rev. A \textbf{111}, 012211 (2025).

\bibitem{Leibfried2003}
D. Leibfried et al.,
Rev. Mod. Phys. \textbf{75}, 281 (2003).

\end{thebibliography}
\end{document}